\documentclass{article}

\usepackage[preprint]{neurips_2025}

\usepackage[utf8]{inputenc} 
\usepackage[T1]{fontenc}    
\usepackage{hyperref}       
\usepackage{url}            
\usepackage{booktabs}       
\usepackage{amsfonts}       
\usepackage{nicefrac}       
\usepackage{microtype}      
\usepackage{xcolor}         
\usepackage{natbib}
\usepackage{thmtools} 

\usepackage[utf8]{inputenc}
\usepackage{fullpage}
\usepackage{amsmath}
\DeclareMathOperator*{\argmax}{arg\,max}

\usepackage{amsthm}
\usepackage{mathtools}
\usepackage{bbm}
\usepackage{frcursive}
\usepackage{commath}
\usepackage{mathrsfs}
\usepackage{xcolor}
\usepackage{amssymb}
\usepackage{hyperref}
\usepackage{comment}
\usepackage[UKenglish]{babel}
\usepackage{bbm}
\usepackage{tikz}
\usetikzlibrary{arrows.meta}

\usepackage{subfig}

\newtheorem{theorem}{Theorem}[section]

\newtheorem{definition}[theorem]{Definition}

\newtheorem{remark}[theorem]{Remark}

\title{Simultaneous Best-Response Dynamics in Random Potential Games\thanks{The authors are grateful to Yannick Viossat and Desmond Chan for the useful discussions and insight they provided.}}

\author{%
Galit Ashkenazi-Golan$^1$ \quad Domenico Mergoni Cecchelli$^{1}$ \quad Edward Plumb$^{1}$\\
$^1$London School of Economics and Political Science\\
\texttt{\{d.mergoni, e.plumb, g.ashkenazi-golan\}@lse.ac.uk}
}

\begin{document}

\maketitle

\begin{abstract}
This paper examines the convergence behaviour of simultaneous best-response dynamics in random potential games. We provide a theoretical result showing that, for two-player games with sufficiently many actions, the dynamics converge quickly to a cycle of length two. This cycle lies within the intersection of the neighbourhoods of two distinct Nash equilibria. For three players or more, simulations show that the dynamics converge quickly to a Nash equilibrium with high probability. Furthermore, we show that all these results are robust, in the sense that they hold in non-potential games, provided the players' payoffs are sufficiently correlated.  
We also compare these dynamics to gradient-based learning methods in near-potential games with three players or more, and observe that simultaneous best-response dynamics converge to a Nash equilibrium of comparable payoff substantially faster. 
\end{abstract}

\section{Introduction}\label{Section_Introduction}
Strategic interactions between agents are typically modelled as games, with the \emph{Nash equilibrium} (NE) serving as the central solution concept. However, this concept requires strong assumptions, including player rationality and, in the presence of multiple equilibria, a principled method for equilibrium selection (\cite{aumann1995epistemic},~\cite{harsanyi1988general}). Moreover, computing a NE often requires knowledge of the opponents' payoffs. These requirements are rarely met in practice, where often solutions are instead found heuristically.

The increasing use of learning agents to address complex optimisation problems raises a central question in game theory and multi-agent learning: \textbf{do learning agents interacting repeatedly converge to a NE?} Such agents adapt their strategies through learning rules designed to improve individual rewards, thereby generating a dynamic process over the space of strategy profiles. These dynamics are called learning dynamics or adaptive dynamics.

A particularly intuitive and widely studied class of learning dynamics is the \emph{best-response dynamics} (see, e.g.~\citet{swenson2018best}), in which players update their strategies to their best response against the current strategy profile of their opponents. There are two primary variants. In the \textbf{sequential} variant, players revise their strategies one at a time, either in a fixed order or according to a stochastic rule (\citet{swenson2018best}). In contrast, the \textbf{simultaneous} variant (from now on SBRD) has all players updating their strategies simultaneously, and is the focus of this work. While the sequential variant requires coordination on the update order, the simultaneous version does not. 

Moreover, SBRD is a \emph{deterministic uncoupled dynamic}, meaning that each player updates their strategy based solely on their own payoffs, without knowledge of others’ payoffs or any need for coordination.  While it is known that dynamics of this family do not always converge to NE in general games (\cite{hart2003uncoupled}), this paper investigates the behaviour of SBRD in the specific context of \emph{random potential games}. 

Potential games have been extensively employed to model a variety of strategic environments, including congestion games~\citet{voorneveld1999congestion, sandholm2010population}, Cournot Competition~\citet{dragone2012static}, and they have applications in theoretical computer science~\cite{Nisan_Roughgarden_Tardos_Vazirani_2007}, wireless
communication~\cite{lasaulce2011game} and evolutionary biology~\cite{hofbauer2003evolutionary}, among others. Potential games, introduced by~\citet{rosenthal1973class} and further developed by~\citet{monderer1996potential} and~\citet{voorneveld2000best}, are games in which there exists a common potential function, mapping action profiles to real or ordinal values, such that each player's optimisation aligns with the optimisation of this global function. In other words, all players effectively aim to maximise the same objective.
In order to estimate typical behaviour, we consider random potential games, i.e. we sample randomly the potential game in which the dynamic occurs. 

The simplicity of SBRD, coupled with the broad applicability of potential games, motivates our central question:

\begin{center}
\emph{If all players in a potential game follow a simultaneous best-response rule,} \\
\emph{is the resulting dynamic likely to converge to a Nash equilibrium?}
\end{center}

Our results indicate that this is the case for three or more players. However, interestingly, we prove that this is not the case for two-player games. Moreover, we show that the same phenomena occur if the potential game assumption is weakened.

Our goal is to estimate the probability with which SBRD converges to a NE in random potential games. The model relies on two assumptions:
(i) the values of the potential function are sampled independently for each action profile from a common distribution. We use the normal distribution in our experiments, although our theoretical results only require sampling from any continuous distribution; and
(ii) all players have the same number of actions. This assumption is made purely for notational convenience and computational simplicity, rather than necessity.

Under these conditions, the resulting probability distribution over best-response trajectories is equivalent to that induced by uniformly sampling an ordering of the action profiles ~\cite{collevecchio2024basins}. This equivalence allows us to study convergence behaviour within the broader framework of random ordinal potential games (see again~\citet{collevecchio2024basins} for further discussion).

\paragraph{Paper outline and summary of results}  
We obtain results in three directions. 
\emph{First}, we characterise the limiting behaviour of SBRD in random potential games as the number of actions per player increases. We do so by providing a formal proof for the two-player case and by giving strong numerical evidence for the cases with three or more players. To our knowledge, ours is the first theoretical result of its kind. \emph{Second}, we verify the robustness of our results by numerically testing whether similar behaviour holds in games that are `close' to potential games ---specifically, games with highly correlated payoffs. \emph{Third}, we compare SBRD with the widely used and well-understood (~\citet{zhang2022global}) softmax policy gradient dynamic, examining both convergence rates and the quality of the resulting payoffs. We now elaborate further.


Firstly, in Section~\ref{sec_twoplayers}, we reveal an interesting difference between games with two players and games with at least three players.  We prove, for random potential games with two players, that with high probability SBRD ends up cycling over a cycle of length two, and thus, not converging to a NE. To the best of our knowledge, no theoretical result analysing convergence of SBRD has been obtained before ours. Furthermore, the convergence to the cycle takes place within a constant number of steps, with a small proportion of the action-profiles being played. This two-cycle consists of two action-profiles $(a,b)$ and $(a',b')$ such that both $(a, b')$ and $(a',b)$ are NE. These results are presented in Section~\ref{sec_twoplayers} and experiments in Section~\ref{exp_SBRD_two_player}. For random potential games with at least three players, we find in Section~\ref{exp_SBRD_three_player} that as the number of actions increases, the probability with which SBRD converges to a NE increases to one.

Secondly, throughout Section~\ref{experimental_results}, we numerically test the robustness of our results to the assumption of the game being a potential game. We simulate random games with various levels of correlation for the payoffs of the players, and find that, in the highly-correlated regime, the results obtained for potential games still hold. With this, we provide strong evidence that highly correlated games behave similarly to potential games with respect to SBRD.

Thirdly, also in Section~\ref{exp_comparison}, we compare SBRD to the softmax policy gradient dynamic (SPGD). We choose SPGD as our benchmark due to its desirable combination of properties: it updates in the direction of the best response while introducing smoothness to the learning dynamics, enjoys strong theoretical convergence guarantees, is well-suited for practical model-free implementation, and incorporates inherent exploration. These features have led to the widespread adoption of softmax policy gradient methods, and their variants, in contemporary reinforcement learning (\cite{mei2020global}, ~\cite{klein2023beyond},~\cite{bernasconi2025evolutionary},~\cite{chen2022convergence} and~\cite{shi2019soft}). We observe that, for three or more players near-potential games, SBRD converges significantly quicker to an equilibrium and scales better to large action sets. We also find that, while SPGD tends to converge to equilibria with moderately higher payoffs, the average payoff along the dynamics is higher for SBRD.
The case of three or more players is presented in Section~\ref{experimental_results}.   

In summary, we show that SBRD cycles around two NE in the case of two players, and converges to a NE in the case of three or more players. We show that this happens quickly, and is robust, meaning that the same holds for games with highly correlated payoffs. Hence, SBRD is a quick and highly-rewarding learning method.

\paragraph{Related work}  
Our research is closely related to two branches of research: learning dynamics in potential games, and learning dynamics in games with random payoffs.

\paragraph{Learning dynamics in potential games}
In recent years, learning dynamics in potential games, and their Markovian extensions, have been extensively studied. Convergence guarantees are of interest: for instance,~\citet{sakos2024beating} analyse 
$q$-replicator dynamics,~\citet{heliou2017learning} prove convergence under no-regret learning with the exponential weights algorithm and minimal information, and~\citet{fox2022independent} show convergence for natural policy gradient learning. Other works focus on the complexity of these dynamics:~\citet{leonardos2021global} study projected gradient dynamics,~\citet{cen2022independent} analyse softmax policy gradient descent with entropy regularisation,~\citet{zhang2022global} examine gradient and natural gradient with log-barrier regularisation,~\citet{ding2022independent} consider projected gradient under various informational assumptions, and~\citet{sun2023provably} investigate natural policy gradient descent methods. More recent contributions include~\citet{dong2024convergence}, who analyse a variant of the Frank-Wolfe algorithm, and~\citet{alatur2024independent}, who study independent policy mirror descent.

\paragraph{Learning dynamics in games with random payoffs}
The behaviour of learning dynamics in games with randomly generated payoffs has been the subject of increasing interest. In the two-player setting,~\citet{galla2013complex} show that experience-weighted attraction learning can lead to a range of outcomes, from convergence to fixed points to complex chaotic behaviour.~\cite{chan2025asymptotic} generalises these results in the many player limit. Assuming the ability for players to coordinate, ~\citet{mimun2024best} demonstrate that, under payoff correlation and a growing number of actions, sequential best-response dynamics converge to a pure NE with high probability. In a similar setting,~\citet{collevecchio2024basins} study two-player random potential games and show that the basin of attraction of each equilibrium is effectively determined by the identity of the player that first updates their strategy.

In games with many players or actions, structural properties of the dynamics are nuanced.~\citet{amiet2021pure} examine large-player games where each player has two actions and payoffs are randomly drawn with a small probability of ties; they show that sequential best-response dynamics typically reach a pure NE as the number of players grows.~\citet{amiet2021better} contrast best and better-response dynamics in two-player games with many actions, finding that while better-response dynamics (with randomly selected updating players) reliably converge to equilibrium when one exists, best-response dynamics tend to enter cycles. This sensitivity to update rules is further emphasised by~\citet{heinrich2023best}, who show that sequential best-response dynamics converge only under random turn-taking; cyclic update orders generally fail to reach equilibrium. Finally,~\citet{johnston2023game} prove that in large random games, any non-equilibrium action profile can be connected via a best-response path to a pure equilibrium, if one exists, with high probability as the action space grows.

\section{The setup}

An $n$-player normal-form game is a triple $( N, (A_i)_{i\in N}, (u_i)_{i\in N})$, where $N= \{1,\dots,n\}$ is a finite set of players, each $A_i$ is the finite action set of player $i$, and $u_i:\prod_{j\in N}A_j \to \mathbb{R}$ is the payoff function of player $i$. For ease of exposition, we assume that all players have the same number $m$ of actions, i.e. for all players $i \in N$ and some $m \in \mathbb{N}$, we have $|A_i| = m$. Let $A = \prod_{i \in N} A_i$ denote the set of action profiles, and, for $i\in N$, let $A_{-i}:=\prod_{j\in N\setminus \{i\}}A_j$. 

Players may randomise over their actions by playing a strategy $x_i\in \Delta(A_i)$, where $\Delta(A_i)$ is the simplex over the set $A_i$. It is standard to extend the payoff function $u_i$ to strategy profiles. And so, for $x=(x_1,\dots,x_n)\in\prod_{i\in N}\Delta(A_i)$, the expected payoff to player $i$ is
\begin{equation}
u_i(x)=\sum_{a\in A}u_i(a) \prod_{j=1}^n x_{j,a_j}\, .    
\end{equation}

With a slight abuse of notation, we sometimes write $a=(a_i,a_{-i})$ and $x = (x_i,x_{-i})$ to denote the combination of player $i$'s action or strategy with the actions or strategies of their opponents. A strategy profile $x^*\in \prod_{i\in N}\Delta(A_i)$ is a NE if there are no profitable unilateral deviations from $x^*$, namely, if for any player $i\in N$ and any strategy $x_i\in \Delta(A_i)$ of this player, it holds that $u_i(x^*_i,x^*_{-i})\geq u_i(x_i, x^*_{-i})$. A strategy profile $x$ is \emph{pure} if each player plays one action with probability $1$. In this case, we often refer to it as an \emph{action profile}.

\subsection{Potential Games}
A game is a potential game if there exists a single function $\Psi\colon A\to\mathbb{R}$ (`the potential') that captures the players' incentives. Formally, 

\begin{definition}
A normal-form game $(N,(A_i)_{i \in N},(u_i)_{i \in N})$ is a potential game if there is a function $\Psi: A\to\mathbb{R}$ such that for each player $i\in N$ there is a constant $c_i\in \mathbb{R}$ such that, for every $a\in A$, we have,
    \begin{align}
    u_i(a) = \Psi(a) + c_i\, .
\end{align}
When all $c_i$ are equal to zero, we simply write $( N,(A_i)_{i \in N},\Psi)$.

\end{definition}
This is equivalent to the classical definition of a potential game (see~\cite{monderer1996potential}).

Thus, a change in player $i$’s payoff from switching actions exactly equals the change in the global potential. Consequently, in a potential game, an action profile is a NE if and only if it is a local maximum of the potential function $\Psi$. 

Without loss of generality, here and in the following we assume all the $c_i$ are equal to $0$. This is not restrictive in our setting as in SBRD players only considers pairwise comparisons of rewards. 

\subsubsection{Random Potential Games}
To study typical behaviour, we introduce the notion of random potential game with $n$ players and $m$ actions.

\begin{definition}
Let $F$ be a continuous real-valued distribution, and $n$ and $m$ positive integers. An $n$-player $m$-actions $F$-random potential game is a potential game $G=(N, (A_i)_{i\in N}, \Psi)$ in which $|N|=n$, and $|A_i|=m$, and moreover we have that for each $a\in A$, the value $\Psi(a)$ is sampled independently at random from $F$.

When $N, A$ and $F$ are clear from context, we just refer to $G$ as a random potential game.
\end{definition}

\subsection{The simultaneous best response algorithm}
One of the simplest learning dynamics is the simultaneous best response dynamic (SBRD). Given a game, starting from an initial action profile $a^0 \in A$, SBRD proceeds as follows: at each round $t \geq 1$ every player $i \in N$ myopically best-responds to the previous action profile $a^{t-1}$. Formally,
\begin{align}
    a_i^t =  \argmax_{a_i \in A_i} u_i(a_i, a_{-i}^{t-1})\, .
\end{align}
If, at some time $t$, we have $a^t = a^{t+1}$, then every player must be playing a best-response to their opponents' strategies, which means $a^t$ is a NE. 

We can assume without loss of generality that $a^0$ is some arbitrary fixed action profile, up to reordering. Once $a^0$ is fixed, since best-response update depends only on the realised potential function $\Psi$, the sequence $(a^t)_{t \geq 0}$ is a random process.

\section{Results}
\label{SBRD_theory}

In this section, we present our main findings on the convergence of SBRD in random potential games. We begin by establishing a theoretical result for two-player games. We show that, for large enough number of actions, SBRD almost surely reaches a two-cycle in a constant number of steps. This two-cycle consists of two action-profiles $(a,b)$ and $(a',b')$ such that  $(a, b')$ and $(a',b)$ are both NE.  We then consider the case of three players or more. Here, we demonstrate via simulations that SBRD converges to a pure NE with probability tending to one as $A$ tends to infinity. 

\subsection{Two Players}\label{sec_twoplayers}

Our main theoretical result is that, in two-player games with sufficiently large action sets, SBRD almost surely converges to a two-cycle in a constant number of steps. 

\begin{restatable}{theorem}{thmMain}\label{thm:Main}
Let $\varepsilon\in (0,1)$,  $F$ be a continuous real distribution, and $G$ be a two-player $m$-actions $F$-random potential game. If $m$ is large enough, then SBRD converges to a two-cycle in at most $\tfrac{\log\varepsilon}{\log(3/4)}$ steps with probability at least $1-\varepsilon$.
\end{restatable}

The main steps of the proof are explained below. All lemmas are proved in Appendix~\ref{TheoreticalAppendix}. The proof of Theorem~\ref{thm:Main} works by comparing the SBRD to another dynamic that converges to a two-cycle, and showing that these two processes coincide up to the termination time with high probability. 

We view the SBRD as a random process over the set of action profiles, where only the payoffs needed are sampled at each time. In this sense, the SBRD for two players proceeds as follows:
\begin{itemize}
    \item At period $0$, the initial action profile $(a^0,b^0) \in A$ is arbitrarily chosen. Also, the following payoffs are sampled (i.i.d. from $F$): $\Psi(a^0,b^0)$, $\Psi(a, b^0)$ for $a\in A_1\setminus \{a^0\}$, and $\Psi(a^0,b)$ for $b\in A_2\setminus\{b^0\}$.
    \item At period $1$, the current action profile is $(a^1,b^1) \in A$ where $a^1:=\argmax_{a\in A_1}\Psi(a, b^0)$, $b^1:=\argmax_{b\in A_2}\Psi(a^0, b)$. As the realised potential values are drawn from a continuous distribution, ties occur with probability zero, and best responses are almost surely unique. Furthermore, the following payoffs are sampled independently from $F$ (if they have not already been sampled):  $\Psi(a^1,b^1)$, $\Psi(a, b^1)$ for $a\in A_1\setminus \{a^1\}$, and $\Psi(a^1,b)$ for $b\in A_2\setminus\{b^1\}$.
    \item In general, at period $t$, the current action profile is $(a^t,b^t) \in A$ where $a^t:=\argmax_{a\in A_1}\Psi(a, b^{t-1})$, and $b^{t}:=\argmax_{b\in A_2}\Psi(a^{t-1}, b)$. Additionally, the following payoffs are sampled independently from $F$ (if they have not already been sampled before):  $\Psi(a^t,b^t)$, $\Psi(a, b^t)$ for $a\in A_1\setminus \{a^t\}$, and $\Psi(a^t,b)$ for $b\in A_2\setminus\{b^t\}$.
    \item This process terminates when there is a repetition, i.e. if at some time $T$ there exists some earlier time $s < T$ such that $(a^T,b^T) = (a^s,b^s)$, then the process terminates at time $T$ in a cycle of length $T-s$. Since the action space is finite, the process must eventually cycle and thus terminate.
\end{itemize}

The first step of our proof is to observe that no cycle of length greater than two can occur. 

\begin{restatable}{lemma}{lemmaTwoCycle}\label{lem:lemmaTwoCycle}
With probability one, the SBRD process terminates at a cycle of length one or two.
\end{restatable}
We can further characterise the two-cycle as follows:
\begin{remark}
    Suppose that the SBRD process does not converge to a NE. By Lemma~\ref{lem:lemmaTwoCycle}, there exists some time $T$, such that $(a^{T-2},b^{T-2}) = (a^T,b^T)$. Consider the two action profiles $(a^{T-1},b^T)$ and $(a^T,b^{T-1})$. As $b^T=b^{T-2}$, we have that $a^{T-1}$ is a best response of player one to $b^T$, and clearly $b^T$ is a best-response of player two to $a^{T-1}$. Hence, $(a^{T-1},b^T)$ is a NE. Similarly, $(a^T,b^{T-1})$ is also a NE. 
\end{remark}

The rest of the proof works by comparing the SBRD with a restricted version of our dynamic, which we call the Independent Dynamic (INDD).
In INDD, at each time $t$, players do not necessarily move to the current best response but rather select the best response amongst the actions they have not yet played, or the action they played in the previous period. While counter-intuitive, because $m$ is large and INDD quickly converges, the set of actions excluded is insignificant compared to the whole set of available actions and therefore the dynamics behave in the same way with high probability.

The reason we consider INDD is that, in this dynamic, at each time $t$, each player's next action is chosen as the maximiser of a set of potential values that are either independent of the history of the process or whose dependence can be carefully controlled. In contrast, under SBRD, any previously sampled payoff that was not the maximiser at the time it was observed becomes less likely to be the maximiser at a later time. This introduces a form of path dependence, thereby breaking the independence structure of the process.

Formally, the INDD is defined as follows.
\begin{itemize}
    \item At time $0$, the initial action profile is $(a^0,b^0)$, and the following payoffs are sampled (i.i.d. from $F$): \[
    \bigl\{\Psi(a,b^0) : a\in A_1\setminus\{a^0\}\bigr\}
    \quad\text{and}\quad
    \bigl\{\Psi(a^0,b) : b\in A_2\setminus\{b^0\}\bigr\}\, .
  \] Note that the value $\Psi(a^0,b^0)$ is not sampled.
  \item At time $1$, the action profile is $(a^1,b^1)$ where: \begin{equation*}
    a^1 := \argmax_{a\in A_1\setminus\{a^0\}} \Psi\bigl(a,b^0\bigr)\hspace{1cm}\text{and}\hspace{1cm} 
    b^1 := \argmax_{b\in A_2\setminus\{b^0\}} \Psi\bigl(a^0,b\bigr)\, .
  \end{equation*}
  Furthermore, all payoffs of the form $\Psi(a,b^{1})$ and $\Psi(a^{1},b)$ that are not known yet are sampled, besides $\Psi(a^{1}, b^{1})$. Note that the set of payoffs for player one that need to be sampled is $R_1^{1}:=\{\Psi(a,b^{1}), a\not\in \{a^\tau, \tau<1\}\}= \{\Psi(a,b^{1}), a \neq a^0\}$ and likewise for player two it is $R_2^{1}:=\{\Psi(a^{1},b), b\not\in \{b^\tau, \tau<1\}\} = \{\Psi(a^{1},b), b \neq b^0\} $. 
    \item At time $t\geq 2$, the action profile is $(a^t,b^t)$ where \begin{align}
        a^t \;&=\; \argmax_{a\in \{a^{t-2}\} \cup (A_1 \setminus  \{a^\tau :  \tau<t\})} \Psi(a,b^{t-1})\, , \\
            b^t \;&=\; \argmax_{b\in \{b^{t-2}\} \cup (A_2 \setminus  \{b^\tau :  \tau<t\})} \Psi(a^{t-1},b)\, .
    \end{align}
    Additionally, all payoffs of the form $\Psi(a,b^{t})$ and $\Psi(a^{t},b)$ that are not known yet are sampled, besides $\Psi(a^{t}, b^{t})$. The set of payoffs for player one that need to be sampled is $R_1^{t}:=\{\Psi(a,b^{t}), a\not\in \{a^\tau, \tau< t\}\}$ and likewise for player two it is $R_2^{t}:=\{\Psi(a^{t},b), b\not\in \{b^\tau, \tau< t\}\}$. 
    \item We define this process to terminate when there is a repetition, i.e. if at some time $T$ there exists some earlier time $s < T$ such that $(a^T,b^T) = (a^s,b^s)$. Then we say that the process terminates at time $T$ in a cycle of length $T-s$. Since the action space is finite, the process must eventually terminate. 
\end{itemize}

In formal statements, we refer to this process as a two-player $m$-actions $F$-INDD.

Since, at time $t$, each player can only play the action that was played at time $t-2$ or one of the actions that they have not played before, the only cycles that can occur are of length two. As the set of action profiles is finite, INDD must cycle, and thus INDD must converge to a cycle of length two. 

The dynamics INDD and SBRD are different only if in SBRD one of the players plays at time $t$ an action that they already played at time $s$ with $s\neq t-2$.

We argue that this occurs with small probability. To this end, we prove first that the INDD process terminates quickly with a high probability.

\begin{restatable}{lemma}{lemmaINDDTermination}\label{lem:INDD-termination}
Let $\varepsilon\in (0,1)$, and $F$ a continuous real-valued distribution. Let us consider a two-player $m$-actions $F$-INDD. If $m$ is large enough, then the probability that the INDD process has not terminated by period $\tfrac{\log\varepsilon}{\log(3/4)}$ is at most $\varepsilon$. 
\end{restatable}

We then show that the probability that INDD and SBRD differ at any step tends to zero as $m\to\infty$.  
A difference between SBRD and INDD can only appear if, for some time $t$, the best response of player one to $b^{t-1}$ is either $a^{t-1}$ or  $a^{\tau}$ for  some $\tau<t-2$ or the analogous happens for player two. The following two lemmas bound the probability of any of these events occurring.

\begin{restatable}{lemma}{lemmaOldActions}\label{lem:old‐actions}
Let $\varepsilon\in (0,1)$, and let $T$ be a positive integer. Consider a two-player $m$-actions $F$-SBRD. If $m$ is large enough, with probability at least $1-\tfrac{\varepsilon}{2}$ there is no $t\leq T$ with either  $a^t\in \{a^0,\dots{}, a^{t-3}\}$ or $b^t\in \{b^0,\dots{}, b^{t-3}\}$.

\end{restatable}

\begin{restatable}{lemma}{lemmaLastProfile}\label{lem:last‐profile}
Let $\varepsilon\in (0,1)$, and let $T$ be a positive integer. Consider a two-player $m$-actions $F$-SBRD. If $m$ is large enough, with probability at least $1-\tfrac{\varepsilon}{2}$ there is no $t\leq T$ with either  $a^t=a^{t-1}$ or $b^t=b^{t-1}$.
\end{restatable}

Combining Lemmas~\ref{lem:old‐actions} and~\ref{lem:last‐profile}, we conclude that for any $\varepsilon > 0$, for any fixed $T$, there exists $\bar m$ such that whenever $m\ge\bar m$,
\begin{align}
\mathbb{P}\bigl(\text{INDD and SBRD coincide up to time }T\bigr)
\;\ge\;1- \varepsilon\, .
\end{align}
Using this result, and that INDD converges quickly to a two-cycle, we obtain Theorem~\ref{thm:Main}.

\section{Experimental Results}\label{experimental_results}

We run extensive simulations for random potential and near-potential games with up to four players.
\paragraph{Key findings.}  (i) In~\ref{exp_SBRD_two_player}, we show that the behaviour proved in Theorem~\ref{thm:Main} persists even in games where player payoffs are highly correlated but not identical. 
(ii) In~\ref{exp_SBRD_three_player} we show that, in the three-player settings, SBRD converges to a Nash equilibrium quickly and with high probability. 
(iii) In~\ref{exp_comparison}, we give evidence that SBRD is considerably faster than SPGD, while obtaining comparable rewards.
\paragraph{Technical details.}  All experiments were executed locally on an Apple MacBook Air with M3 chip with 16~GB RAM with no use of GPU. Code and data are publicly available in supplementary material. 
Metrics on continuous-valued variables are plotted with $\pm 2$ standard errors (SE); binomial metrics are presented with $99.5\%$ Clopper--Pearson confidence intervals.

\subsection{Numerical Setup}\label{Numerical_setup}
Let $n$ denote the number of players, $m$ the number of actions, $s$ the number of samples, and $\lambda \in [0,1]$ the correlation parameter. For each experiment, we generate $s$ independent $n$-player $m$-action games. For each action profile $a \in A$, the payoff $u_i(a)$ is drawn from a standard normal distribution with pairwise correlation $\lambda$ between any two players $i \ne j$. Samples are taken independently for each $a$. As in~\cite{galla2013complex}, we argue that this is the natural choice because, given the first and second degree moments, it is entropy maximising. 

We vary $\lambda$ over \texttt{[0.05 * i for i in range(21)]} to cover the full $[0,1]$ range, and over \texttt{[0.85 + 0.025 * i for i in range(7)]} to test robustness to the potential game assumption. While a finer discretisation is possible, we find these values sufficient to illustrate the trends. The choice of $m$ and $n$ are described for each experiment.

\subsection{SBRD in two-player games}\label{exp_SBRD_two_player}
\begin{figure}%
    \centering
    \subfloat[\centering Probability of converging to a two-cycle.]{{\includegraphics[width=6.7cm]{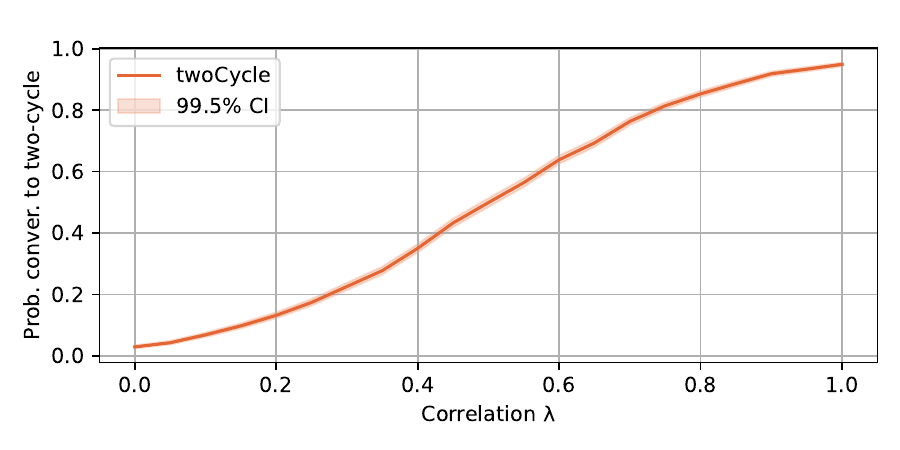}\label{S1twoCycles} }}%
    \quad
    \subfloat[\centering Number of steps to convergence.]{{\includegraphics[width=6.7cm]{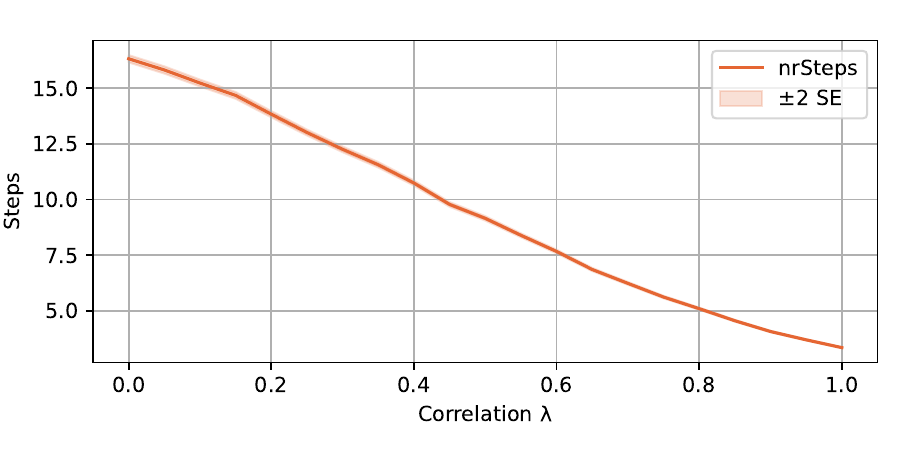}\label{S1steps} }}%
    \caption{SBRD in a two-player $50$-actions game. 10000 samples were drawn. Runtime: $22$ seconds.}%
    \label{fig:twoPlayer}%
\end{figure}
Our main findings, supported by Figure~\ref{fig:twoPlayer}, are to support and show robustness of Theorem~\ref{thm:Main}. Figure~\ref{S1twoCycles} illustrates the findings regarding two-player $50$-actions games, and shows that, for high values of correlation $\lambda$, SBRD is likely to quickly converge to a two-cycle. For $\lambda=1$ we rediscover the statement of our Theorem ~\ref{thm:Main}. Figure~\ref{S1steps} also shows that the number of steps to convergence diminishes drastically with higher values of $\lambda$.

We address the assumption $m=50$ in Appendix~\ref{sec:robustnessTwoPlayer}, where we show that the same behaviour occurs for $m=500$ (and therefore it is not reductive to assume $m=50$ in this case).

\subsection{SBRD in three (or more)-player Games}\label{exp_SBRD_three_player}
\begin{figure}%
    \centering
    \subfloat[\centering Probability of converging to a NE.]{{\includegraphics[width=6.7cm]{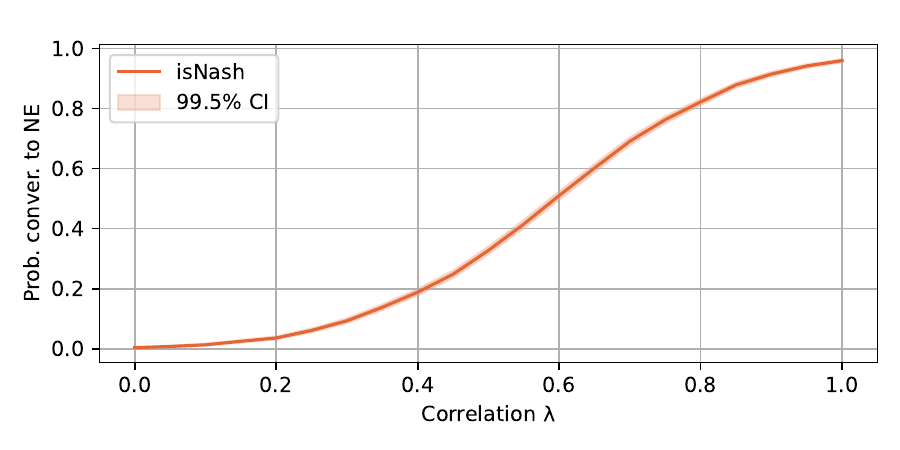}\label{S2NE} }}%
    \quad
    \subfloat[\centering Number of steps to convergence.]{{\includegraphics[width=6.7cm]{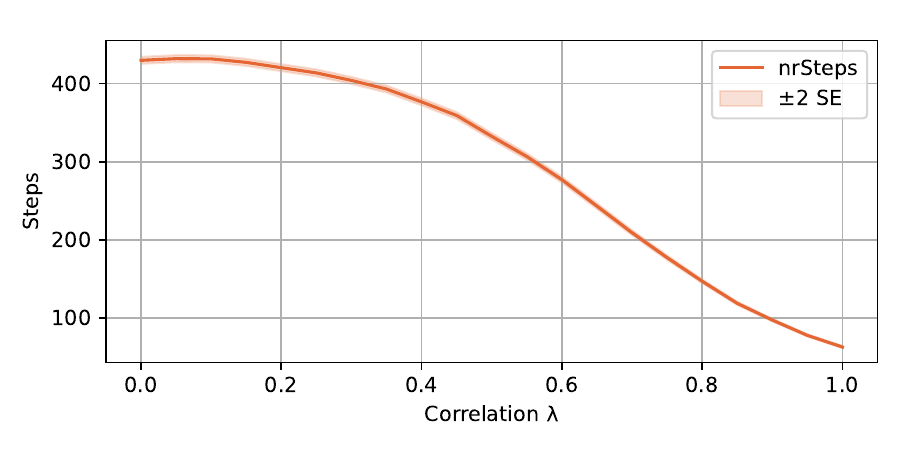}\label{S2Steps} }}%
    \caption{SBRD in a three-player $50$-actions game. $1000$ samples were drawn. Runtime: $20$ minutes.}%
    \label{fig:threePlayer}%
\end{figure}
Figure~\ref{S2NE} provides strong empirical evidence that, in contrast to the two-player case, SBRD is likely to converge to a NE in three-player random potential games. This behaviour is not only prevalent in potential games, but also persists in games with sufficiently high payoff correlation $\lambda$. 
As for the two-player case, Figure~\ref{S2Steps} shows that convergence happens in a number of steps that diminishes for higher values of correlation $\lambda$.

As before, we postpone to Appendix~\ref{sec:robustnessThreeActions} to show that the assumption $m=50$ is not reductive, and that similar behaviour occurs for $m=100$.

Also in Appendix~\ref{sec:robustnessThreeToPlayers} we address the case with four players, showing for the case $n=4$, $m=50$ that the same behaviour occurs in this setting as well. We conjecture that this trend extends to games with more than four players. However, we did not pursue this direction further, as we believe that the three- and four-player cases already provide strong evidence. 

\subsection{Comparison of SBRD and SPGD in three-player near-potential games}\label{exp_comparison}

We now consider near-potential games with $\lambda \geq 0.85$ and compare SBRD with SPGD. As previously discussed in the paper outline, we selected SPGD as a natural baseline due to its smooth best-response updates, convergence guarantees, and model-free applicability. 

\begin{figure}%
    \centering
    \subfloat[\centering Time to convergence in SBRD and SPGD.]{{\includegraphics[width=6.7cm]{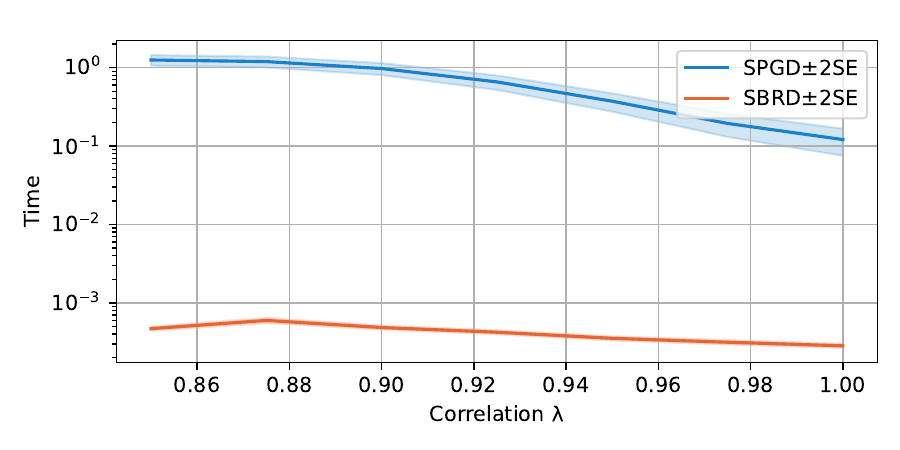}\label{S3Time} }}%
    \quad
    \subfloat[\centering Final value of equilibrium.]{{\includegraphics[width=6.7cm]{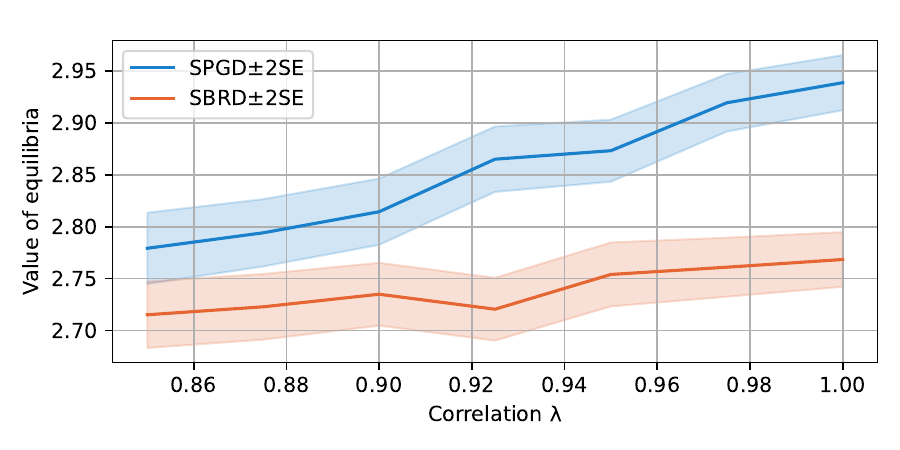}\label{S3ValFin} }}%
    \caption{Comparison of SPGD and SBRD in a three-player $50$-actions game. $1000$ samples were drawn. Runtime: $80$ minutes.}%
    \label{fig:comparison}%
\end{figure}

As in the previous section, we focus on three-player games with $50$ actions. Figure~\ref{S3Time} shows that SBRD converges drastically faster than SPGD. Empirically,  the time from start to convergence under SBRD is roughly three orders of magnitude lower than for SPGD. 

In terms of achieved payoffs, SPGD tends to attain marginally higher equilibrium payoffs, but the difference remains small (Figure~\ref{S3ValFin}). Crucially, as shown in Section~\ref{appendix_Complement_comparison}, SPGD often requires several thousand iterations to converge, and during its trajectory, the average payoff is much lower. On the other hand, as shown above in Figure~\ref{S2Steps}, SBRD consistently converges in under $100$ iterations when $\lambda \geq 0.9$. In Section~\ref{appendix_Complement_comparison} we quantify precisely the number of steps needed on average for SPGD to converge, and the average payoff of SPGD compared to the equilibrium value attained by SBRD. This speed advantage and the payoff comparison persist in the $100$-actions setting as shown in Section~\ref{appendix_Robustness_comparison}. We thus claim that SBRD provides a favourable trade-off, especially in online settings, delivering comparable payoffs at a small fraction of the computational cost.

\section{Discussion and Limitations}\label{discussionAndLimitations}

In this work, we have analysed Simultaneous Best Response Dynamics (SBRD) in the setting of random potential games. In contrast to sequential best response dynamics, SBRD requires no centralised coordination on the order of updates: at each round, every player updates their action to a best response against the joint profile of their opponents. This feature causes SBRD to be a more plausible model of strategic adaptation in decentralised multi‐agent systems.

Our findings exhibit an interesting dependence on the number of players. In the two‐player case, SBRD enters a two‐cycle with high probability in games with highly correlated payoffs. In particular, the players alternate between two action profiles involving mismatched actions from two distinct Nash equilibria. Although such oscillatory behaviour prevents convergence, introducing a small random perturbation to each best‐response update would break the cycle and restore convergence to a Nash Equilibrium (NE). By contrast, in games with three or four players, our simulations suggest that SBRD tends to converge very quickly to a NE. Moreover, when benchmarking against Softmax Policy Gradient Dynamics (SPGD), we observe that SBRD achieves higher learning‐phase payoffs, even if SPGD tends to perform slightly better in terms of final payoffs at convergence. We conjecture that this also holds for $n$-player potential games where $n \ge 5$. 

Moreover, since best‐response updates depend solely on the ordinal ranking of payoffs, all results carry over to ordinal potential games in the sense of~\cite{monderer1996potential}.  We further demonstrate empirically that our conclusions are robust when the payoff‐correlation assumption is relaxed: games with highly correlated payoffs exhibit the same convergence behaviours.

\paragraph{Assumptions and Limitations} 

Our theoretical analysis focuses on two‐player random potential games, hence we assume a perfectly correlated payoff structure.  Although exact payoff alignment is uncommon in practical settings, the potential‐game framework encompasses a broad class of models (see Section~\ref{Section_Introduction}), and our empirical investigations indicate that the core convergence behaviour persists when payoffs are merely highly, rather than perfectly, correlated.  

All experimental findings are derived from simulations in which payoff entries are drawn from a normal distribution. As previously argued, this is the natural entropy-maximising choice. However, this choice may not capture the diversity of strategic environments; exploring alternative distributions (e.g.\ uniform, heavy‐tailed or bimodal) could reveal new phenomena. 

At present, a rigorous proof of convergence for $n\ge3$ players remains outstanding, and we view the extension of our theoretical guarantees to game with more players as an important avenue for future work.  Likewise, while we benchmark SBRD only against SPGD, other adaptive schemes, such as Q‐learning, replicator dynamics or fictitious play, may exhibit different performance characteristics and merit systematic comparison.

Finally, our model assumes that each player has complete knowledge of their own payoffs and full observability of opponents’ actions.  Relaxing these assumptions to allow for partial observability or payoff estimation through exploration would make the model more realistic, but at the cost of substantially greater analytical complexity.  We defer the study of such extensions to future research.  

We are not aware of ethical implications of our research, as it is an analysis of pre-existing concepts.

\paragraph{Summary} 
The Simultaneous Best-Response Dynamic is a simple yet powerful learning rule, with provable convergence behaviour in two-player potential games and promising empirical performance in potential and near-potential games with more players. Its key limitation is the current gap between numerical conjectures and formal proofs for games with more than two-players. Addressing this challenge will deepen our theoretical understanding and broaden the applicability of SBRD.

\bibliography{br_dynamics_new}
\bibliographystyle{plainnat}

\medskip

\newpage
\appendix

\section{Theoretical Appendix}\label{TheoreticalAppendix}
In this appendix, we provide the proofs of section~\ref{sec_twoplayers}.

\subsection{Observations regarding INDD}\label{AppendixObservations}
We start from some observations regarding INDD:
\begin{itemize}
    \item Since, at time $t$, each player can only play the action that was played at time $t-2$ or one of the actions that they have not played before, the only cycles that can occur are of length two. As the set of action profiles is finite, INDD must cycle, and thus INDD must converge to a cycle of length two. 
    \item At each time $t$, we have $\Psi(a^{t}, b^{t-1}) = \max (\Psi(a^{t-2}, b^{t-1}), \max R_1^{t-1})$ and $\Psi(a^{t-1}, b^{t}) = \max (\Psi(a^{t-1}, b^{t-2}), \max R_2^{t-1})$.
    \item Once either player repeats their previous but one action, there is always one player repeating their previous but one action, in an alternating manner.  For example, suppose that at period $t$, Player 1 chooses $a^t = a^{t-2}$, then at period $t+1$ Player 2 chooses $b^{t+1} = b^{t-1}$.
\end{itemize}

We are now ready to follow the proofs.

\subsection{Proof of Lemma~\ref{lem:lemmaTwoCycle}}

\begin{proof}
    Define two sequences $(M_\ell)$ and $(N_\ell)$ for $\ell=1,2,\dots$ by:
    \begin{align}
        M_\ell &= \begin{cases}
            \Psi(a^{\ell},b^{\ell -1}) & \text{if } \ell \text{ is odd,}  \\
            \Psi(a^{\ell -1},b^{\ell}) & \text{if } \ell  \text{ is even.}
        \end{cases} \\
        N_\ell &= \begin{cases}
            \Psi(a^{\ell -1},b^{\ell}) & \text{if } \ell \text{ is odd, } \\
            \Psi(a^{\ell},b^{\ell-1}) & \text{if } \ell \text{ is even. }
        \end{cases}
    \end{align}
    
Hence,
\begin{align}
    (M_\ell)_{\ell\ge1} &= (\Psi(a^1,b^0), \Psi(a^1,b^2), \Psi(a^3,b^2),\dots{}) \\
    (N_\ell)_{\ell\ge1} &= (\Psi(a^0,b^1), \Psi(
    a^2,b^1), \Psi(a^2,b^3),\dots{})
\end{align}

Observe that each transition $M_\ell\to M_{\ell+1}$ is a best‐response transition by one of the players, so almost surely $M_{\ell+1} > M_\ell$, unless the opponent’s action does not change (i.e. $b^\ell=b^{\ell+2}$ when $\ell$ is even, or $a^\ell=a^{\ell+2}$ when $\ell$ is odd).  But note that, if $b^\ell=b^{\ell+2}$ for some even $\ell$, then one obtains
\begin{align}
    a^{\ell+1}=a^{\ell+3}, \quad b^{\ell+2}=b^{\ell+4}, \quad \dots
\end{align}
and the same holds if $a^\ell=a^{\ell+2}$ when $\ell$ is odd, which makes $(M_\ell)$ a one-cycle.  
So, with probability one, either $M_1 < M_2 < \dots$ or $(M_\ell)$ is a one-cycle. 

An identical argument applies to $(N_\ell)$.  But then both players’ actions have period at most 2, and so $T \leq 2$.  
Therefore, with probability one, no cycle of length greater than 2 can occur.

Since the space is finite, neither of the sequences can increase indefinitely, and therefore will eventually cycle.
\end{proof}

\subsection{Proof of Lemma~\ref{lem:INDD-termination}}

\begin{proof}

We split the argument into two parts. 

Fix $\varepsilon >0$ and a horizon $T \in \mathbb{N}$. The first part is to show that for all periods $t \le T$, the probability that at least one player repeats their action from period $t-2$ is at least $1/2$, provided $m$ large enough. Equivalently, at each $t$, at least one of the events
\begin{align}
E^{t-1}_1 &: \Psi\bigl(a^{t-2},b^{t-1}\bigr)\;>\;\max R_1^{\,t-1}\,,\\
E^{t-1}_2 &: \Psi\bigl(a^{t-1},b^{t-2}\bigr)\;>\;\max R_2^{\,t-1}\,,
\end{align}
occurs with probability at least $1/2$. 

The second part is to show that for large enough $m$, if exactly one of these events takes place, then from that period on, the probability that both events happen is at least $\frac{1}{2}$. 

Once these parts are done, for $m$ large enough to satisfy both conditions, it holds that for $T\ge\frac{\log\varepsilon}{\log(3/4)}$, the probability that INDD lasts more than $T$ periods is less than $\varepsilon$.

For the first part, we focus on path of comparisons through the space of action profiles, where:
\begin{itemize}
    \item Player 1 chooses $a^1$ to maximise $\Psi(a,b^0)$. 
    \item Player 2 compares $\Psi(a^1,b^0)$ to the newly revealed values to choose $b^2$.
    \item Player 1 compares $\Psi(a^1,b^2)$ to the newly revealed values to choose $a^3$.
    \item And so on \dots{}
\end{itemize}

The choice of starting with Player 1 is arbitrary. There is an equivalent path that begins with Player 2 choosing $b^1$.

In the first step of the path, Player 1 selects
\begin{align}
a^1 \;=\;\arg\max_{a\neq a^0}\Psi(a,b^0)\,,
\end{align}
so that
\begin{align}
\Psi(a^1,b^0)
=\max\bigl\{\Psi(a,b^0):a\in A_1\setminus\{a^0\}\bigr\}\,,
\end{align}
is the maximum of $m-1$ independent draws from $F$.  

In the second step of the path, Player 2 only knows $\Psi(a^1,b^0)$ and draws $m-2$ new payoffs 
\begin{align}
R_2^1=\{\Psi(a^1,b):b\in A_2\setminus\{b^0,b^1\}\}\,.
\end{align}
and they return to playing $b^0$ precisely if 
\begin{align}
 \Psi(a^1,b^0) \;>\; \max R_2^1\,.
\end{align}
By symmetry of i.i.d. sampled from $F$,
\begin{align}
\mathbb{P}(\Psi(a^1,b^0) \;>\; \max R_2^1)
=\frac{m-1}{(m-1) + (m-2)}
>\frac{1}{2}
\quad(\text{for }m \ge 3)\,.
\end{align}
Hence, the event $E^1_2$ occurs with probability at least $1/2$. 

In the event that $E^1_2$ does not occur, then at period 2 Player 2 is playing $b^2$, and $\Psi\bigl(a^{1},b^{2}\bigr)$ is the maximum of $2m-3$ i.i.d. sampled from $F$. Then, $m-3$ new payoffs are randomised, and by the same symmetry argument, we have
\begin{align}
\mathbb{P}(E^2_1) =
\mathbb{P}(\Psi\bigl(a^{1},b^{2}\bigr)\;>\;\max R_1^{\,2})
=\frac{2m-3}{(2m-3)+(m-3)}
>\frac{1}{2}\,.
\end{align}
In general, consider period $t$. If $t$ is odd, then, in the event that $E^1_2, E^2_1, E^3_2,\dots{},E^{t-1}_1$ all did not occur, then Player 1 is playing $a^t$, and $\Psi\bigl(a^{t},b^{t-1}\bigr)$ is the maximum of $\sum_{\tau=1}^{t-1}(m-\tau)=(t-1)m -\frac{t(t-1)}{2} $ i.i.d.\ from $F$. The realisations of $m-t$ variables are observed, and hence:
\begin{align}
\mathbb{P}(E^{t}_2) = 
\mathbb{P}(\Psi(a^t,b^{t-1}) \;>\; \max R_2^t)
=\frac{(t-1)m -\frac{t(t-1)}{2}}{tm -\frac{t(t+1)}{2}}
>\frac{1}{2}\,.
\end{align}
For $t$ even, analogously we can show that $\mathbb{P}(E^{t}_1)  
>\frac{1}{2}$. Then, for $m$ large enough, this holds for all $t<T$.

For the second part, we suppose that at some period $t$ exactly one of the events $E_1^{\,t-1}$ or $E_2^{\,t-1}$ occurs; without loss of generality assume 
\begin{align}
    E_1^{\,t-1}:\;\Psi\bigl(a^{t-2},b^{t-1}\bigr)\;>\;\max R_1^{\,t-1}
\quad\text{and}\quad
\neg E_2^{\,t-1}:\;\Psi\bigl(a^{t-1},b^{t-2}\bigr)\;\le\;\max R_2^{\,t-1}\,.
\end{align}
Then Player 1 re-plays action $a^{t-2}$, so $a^t=a^{t-2}$, while Player 2 plays a new action $b^t\neq b^{t-2}$.  Hence the action profile at time $t$ is 
\[
(a^t,b^t)=\bigl(a^{t-2},b^t\bigr)\,.
\]

We show that the probability that the process terminates in the next period is at least $1/2$, provided $m$ large enough.  

At period $t+1$, Player 1 compares the known value $\Psi(a^{t-1}, b^{t})$ (which is the maximum of at least $m-1$ independent draws from $F$) to the maximum of the newly realised payoffs in $R^t_1$, which contains at most $m-1$ new samples from $F$. Meanwhile, Player 2 compares the known value $\Psi(a^{t}, b^{t-1})=\Psi(a^{t-2}, b^{t-1})$  to no newly generated values (since $a^{t-2}$ was just re-played), and so replays $b^{t-1}$.

Thus at period $t+1$, by the same symmetry argument as before, the probability that Player 1 repeats $a^{t-1}$ again is at least:
\begin{align}
\frac{m-1}{(m-1)+(m-1)}
=\frac12\,.
\end{align}
Hence, with probability at least $1/2$, the action profile $(a^{t-1},b^{t-1})$ is repeated, and so the process terminates at period $t+1$.

Putting the two parts together: choose $m_0$ large enough that in each period $t\le T$ both  
\begin{align}
\mathbb{P}\bigl(E_1^{\,t}\cup E_2^{\,t}\bigr)\ge\frac12
\quad\text{and}\quad
\mathbb{P}\bigl(\text{termination}\mid\text{exactly one of }E_1^{\,t-1},E_2^{\,t-1}\bigr)\ge\frac12\,.
\end{align}
Then the probability the process survives beyond $T$ is bounded above by 
\begin{align}
\bigl(1-\tfrac12\cdot\tfrac12\bigr)^{\!T}
=\bigl(\tfrac34\bigr)^T\,,
\end{align}
and for $T\ge\frac{\log\varepsilon}{\log(3/4)}$ this is at most $\varepsilon$.  

\end{proof}

\subsection{Proof of Lemma~\ref{lem:old‐actions}}

\begin{proof}
Fix any horizon $T$.  For $t=3,\dots{},T$, let $E^1_t$ be the event that
\begin{align}
\max_{t' < t-2} \Psi(a^{t'},b^{t-1})  > \Psi(a^{t-2},b^{t-1})\,.
\end{align}
Using the same argument as in the previous proposition, $\Psi(a^{t-2},b^{t-1})$ must be the maximum of $(t-1)m -\frac{t(t-1)}{2} $ i.i.d samples from $F$. Therefore, if $E^1_t$ occurs then one of the $t-2$ payoffs $\{\Psi(a^{t'},b^{t-1}):t'<t-2\}$ must exceed this maximum.  By symmetry, for each fixed $t$
\begin{align}
\mathbb{P}(E^1_t)\;\le \frac{t-2}{\,(t-1)m -\frac{t(t-1)}{2}\,} \le\;\frac{t-2}{\,m-1\,}\,.
\end{align}
Hence by the union bound,
\begin{align}
\mathbb{P}\Bigl(\bigcup_{t=3}^T E^1_t\Bigr)
\;\le\;\sum_{t=3}^T\frac{t-2}{m-1}
=\frac{\,(T-2)(T-1)/2\,}{m-1}\,,
\end{align}
which can be made below $\varepsilon /4$ by choosing $m$ large.  One can define $E^2_t$ to be the analogous event for player two, and achieve that $\mathbb{P}\Bigl(\bigcup_{t=3}^T E^2_t\Bigr) \leq \varepsilon / 4$ by the same argument. This bounds the probability that SBRD differs from INDD on account of any `old' action–payoff comparison.
\end{proof}

\subsection{Proof of Lemma~\ref{lem:last‐profile}}

\begin{proof}
Again fix horizon $T$.  At each period $t=1,\dots{},T$, SBRD additionally compares the single value $\Psi(a^{t-1},b^{t-1})$ against at least $m-t-1$ fresh samples of the distribution $F$.  By symmetry the chance it is the maximum is
\begin{align}
\frac{1}{(m-t-1)+1}
=\frac{1}{m-t}\,.
\end{align}
Over $T$ periods, a union‐bound gives
\begin{align}
\Pr\bigl(\exists\,t\le T:\text{SBRD uses }\Psi(a^{t-1},b^{t-1})\bigr)
\;\le\;\sum_{t=1}^T\frac{1}{m-T}
=\frac{T}{m-T}\,,
\end{align}
which is below $\varepsilon/2$ for all $m\ge\frac{T(1+\varepsilon/2)}{\varepsilon/2}$.
\end{proof}

\subsection{Proof of Theorem~\ref{thm:Main}}

\begin{proof}
Fix $\varepsilon>0$.  We show that for sufficiently large $m$ three events each occur with probability at least $1-\tfrac\varepsilon3$, and hence by the union bound the SBRD process converges to a 2‑cycle with probability at least $1-\varepsilon$.  

Firstly, by Lemma~\ref{lem:INDD-termination}, there exist an integer $m_{0}$ such that whenever $m\ge m_{0}$ the INDD process terminates by period $T=\frac{\log(\varepsilon/3)}{\log(3/4)}$ with probability at least $1-\tfrac\varepsilon3$.

As mentioned in~\ref{AppendixObservations}, INDD cannot terminate in a 1‑cycle and cannot cycle with length $>2$.  Hence on termination it must enter a 2‑cycle with probability one (and so at least $1-\tfrac\varepsilon3$).

By Lemmas~\ref{lem:old‐actions} and~\ref{lem:last‐profile}, there exists $m_{1}$ such that whenever $m\ge m_{1}$ the probability that INDD and SBRD differ at some period $t\le T$ is at most $\tfrac\varepsilon3$.  Equivalently, with probability at least $1-\tfrac\varepsilon3$ they coincide up to time $T$.

Therefore, if $m\ge \max\{m_{0},m_{1}\}$, then each of the three events has probability at least $1-\tfrac\varepsilon3$, so by the union bound all three occur simultaneously with probability at least  
\begin{align}
    1 - 3\cdot\tfrac\varepsilon3 = 1-\varepsilon\,.
\end{align}
 
In that event, SBRD follows the same path as INDD up to period $T$, INDD terminates in a two‑cycle by $T$, and hence SBRD too converges to that same two‑cycle.  Therefore 
\begin{align}
    \mathbb{P}\bigl(\text{SBRD converges to a two‑cycle by time } T\bigr)\;\ge\;1-\varepsilon\,,
\end{align}
 as required.  
\end{proof}

\section{Experimental Appendix}\label{Experimental_appendix}

\subsection{Robustness to Number of Actions for Section~\ref{exp_SBRD_two_player}}\label{sec:robustnessTwoPlayer}
\begin{figure}%
    \centering
    \subfloat[\centering Probability of converging to a two-cycle.]{{\includegraphics[width=6.7cm]{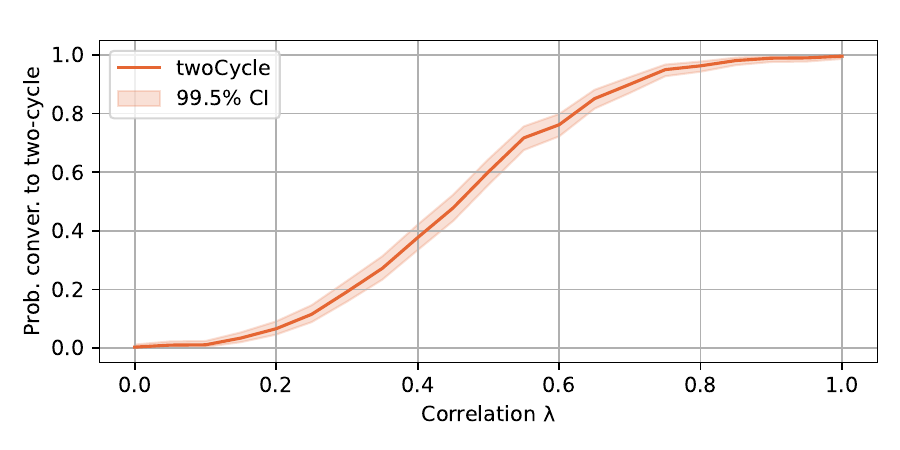}\label{S4twoCycles} }}%
    \quad
    \subfloat[\centering Number of steps to convergence.]{{\includegraphics[width=6.7cm]{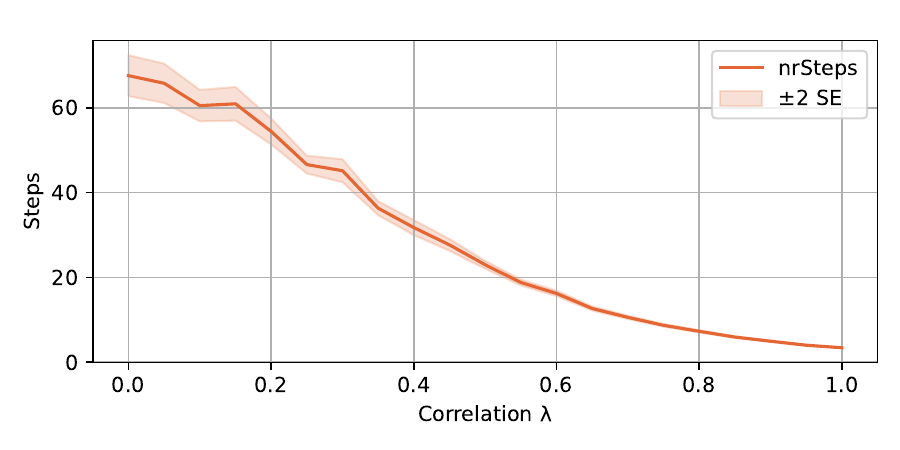}\label{S4Steps} }}%
    \caption{SBRD in a two-player $500$-actions game. $1000$ samples were drawn. Runtime: $127$ seconds}%
    \label{fig:AppendixTwoPlayer}%
\end{figure}

We now complement the experimental findings of Section~\ref{exp_SBRD_two_player} by showing that the results are robust with respect to the number of actions. In particular, Figure~\ref{fig:AppendixTwoPlayer} shows that two-player random games with $500$ actions exhibit the same behaviour as in the $50$-action case: the probability of convergence to a two-cycle varies similarly with $\lambda$, and the number of steps required to converge in highly correlated games remains of the same order of magnitude.

To achieve a convergence probability of at least $90\%$, values of $\lambda \geq 0.9$ were needed for $m = 50$, whereas for $m = 500$, values of $\lambda \geq 0.75$ were sufficient. This suggests that the behaviour predicted by Theorem~\ref{thm:Main} extends to larger games and can emerge even at lower levels of correlation.

These findings strongly support the claim made in Section~\ref{exp_SBRD_two_player} that in two-player highly correlated random games, SBRD quickly converges to a two-cycle. 

The experiment shown in Figure~\ref{fig:AppendixTwoPlayer} ran in $127$ seconds.

\subsection{Robustness to Number of Actions for Section ~\ref{exp_SBRD_three_player}}\label{sec:robustnessThreeActions}

\begin{figure}%
    \centering
    \subfloat[\centering Probability of converging to a NE.]{{\includegraphics[width=6.7cm]{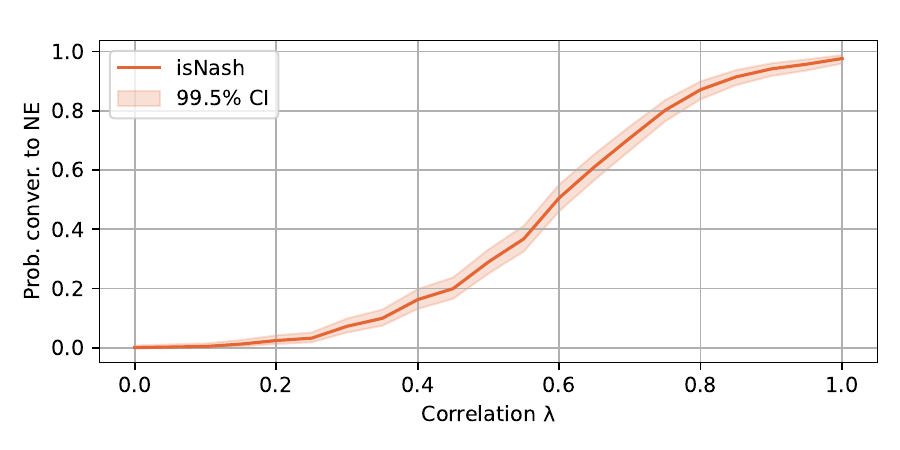}\label{S5NE} }}%
    \quad
    \subfloat[\centering Number of steps to convergence.]{{\includegraphics[width=6.7cm]{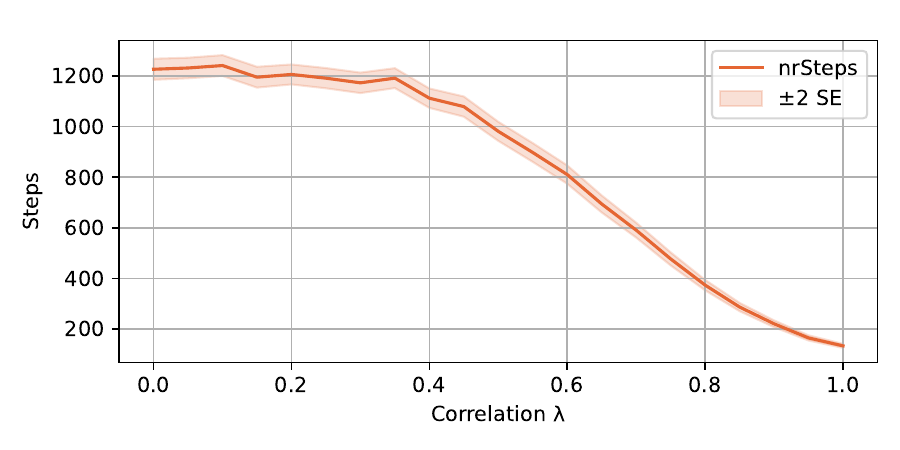}\label{S5Steps} }}%
    \caption{SBRD in a three-player $100$-actions game. $1000$ samples were drawn. Runtime: $14$ minutes.}%
    \label{fig:AppendixThreePlayer}%
\end{figure}

We now show with Figure~\ref{fig:AppendixThreePlayer} that the number of actions does not affect the outcomes reported in Section~\ref{exp_SBRD_three_player}. Specifically, we run experiments on three-player random games with $100$ actions across various levels of correlation $\lambda$. The results closely mirror those observed in the $50$-action case. For high values of $\lambda$, the probability that SBRD converges to a Nash equilibrium approaches one, and this behaviour appears smoothly as correlation increases. In other words, in highly correlated games, SBRD is very likely to converge to a Nash equilibrium.

We also observe that the number of steps required for convergence remains of the same order of magnitude across both settings when $\lambda$ is large.

This new evidence reinforces the claim made in Section~\ref{exp_SBRD_three_player} that in highly correlated three-player games, SBRD tends to quickly converge to a Nash equilibrium.

The experiment shown in Figure~\ref{fig:AppendixThreePlayer} ran in $14$ minutes.

\subsection{Robustness to Number of Players for Section ~\ref{exp_SBRD_three_player}}\label{sec:robustnessThreeToPlayers}
\begin{figure}%
    \centering
    \subfloat[\centering Probability of converging to a NE.]{{\includegraphics[width=6.7cm]{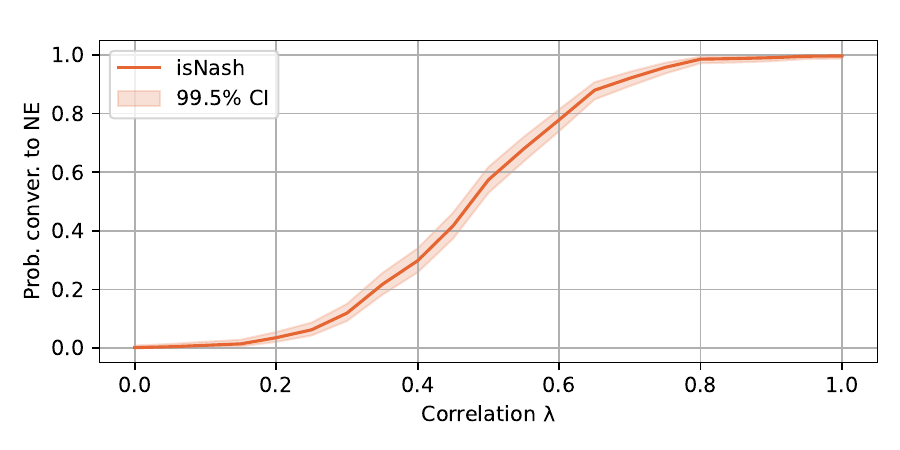}\label{S6NE} }}%
    \quad
    \subfloat[\centering Number of steps to convergence.]{{\includegraphics[width=6.7cm]{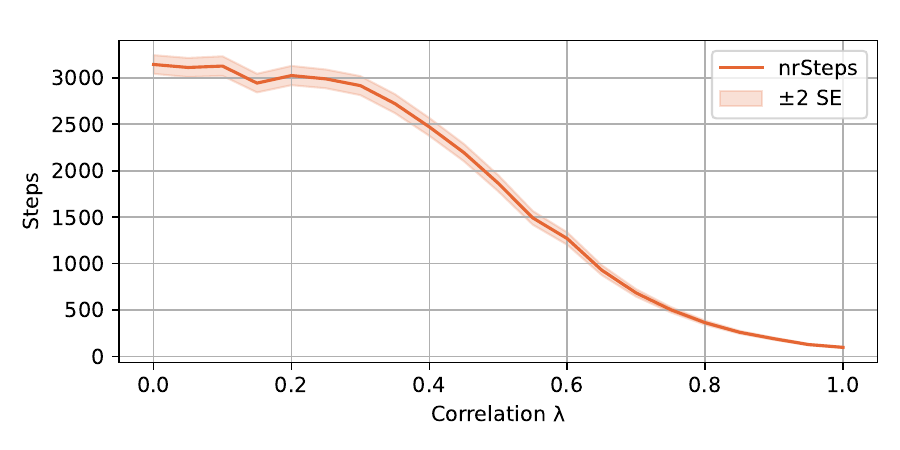} \label{S6Steps}}}%
    \caption{SBRD in a four-player $50$-actions game. $1000$ samples were drawn. Runtime: $142$ minutes.}%
    \label{fig:AppendixthreePlayerfour}%
\end{figure}

Having shown that the number of actions does not influence the behaviour of SBRD across different levels of $\lambda$, we now see if the behaviour is influenced by the number of players. As previously mentioned, we believe that the convergence to a two-cycle (and thus not to a NE) observed in the two-player setting is a special case, and that for games with three or more players and high payoff correlation, SBRD is likely to converge to a Nash equilibrium.

Figure~\ref{fig:AppendixthreePlayerfour} confirms that SBRD behaves in the four-player case as it does in the three-player setting. Specifically, the probability of convergence to a NE is very high for large values of $\lambda$, and the number of steps required to converge decreases sharply as correlation increases.

While we do not experimentally test games with more than four players, nor provide a formal proof, ongoing research is aimed at establishing this behaviour theoretically.

The experiment shown in Figure~\ref{fig:AppendixthreePlayerfour} ran in $142$ minutes.

\subsection{Complement to Section~\ref{exp_comparison}}\label{appendix_Complement_comparison}
\begin{figure}%
    \centering
    \subfloat[\centering Steps to convergence in SBRD and SPGD.]{{\includegraphics[width=6.7cm]{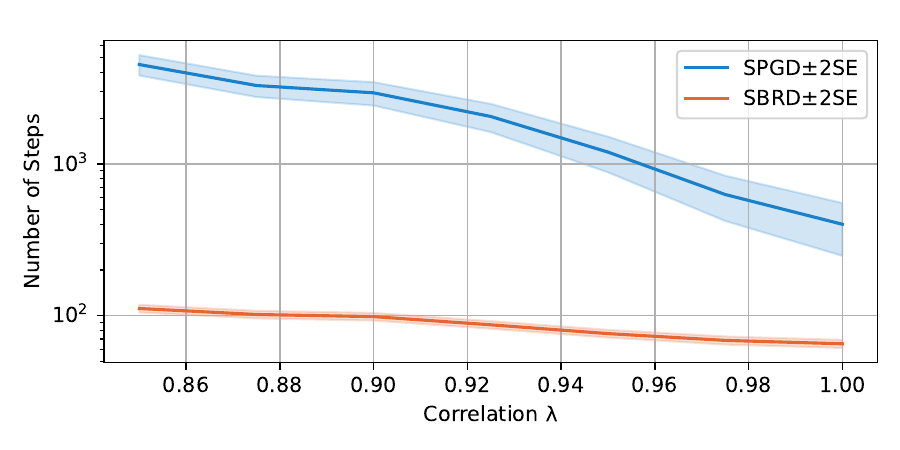}\label{S3Steps} }}%
    \quad
    \subfloat[\centering SBRD equilibrium value vs SPGD average value.]{{\includegraphics[width=6.7cm]{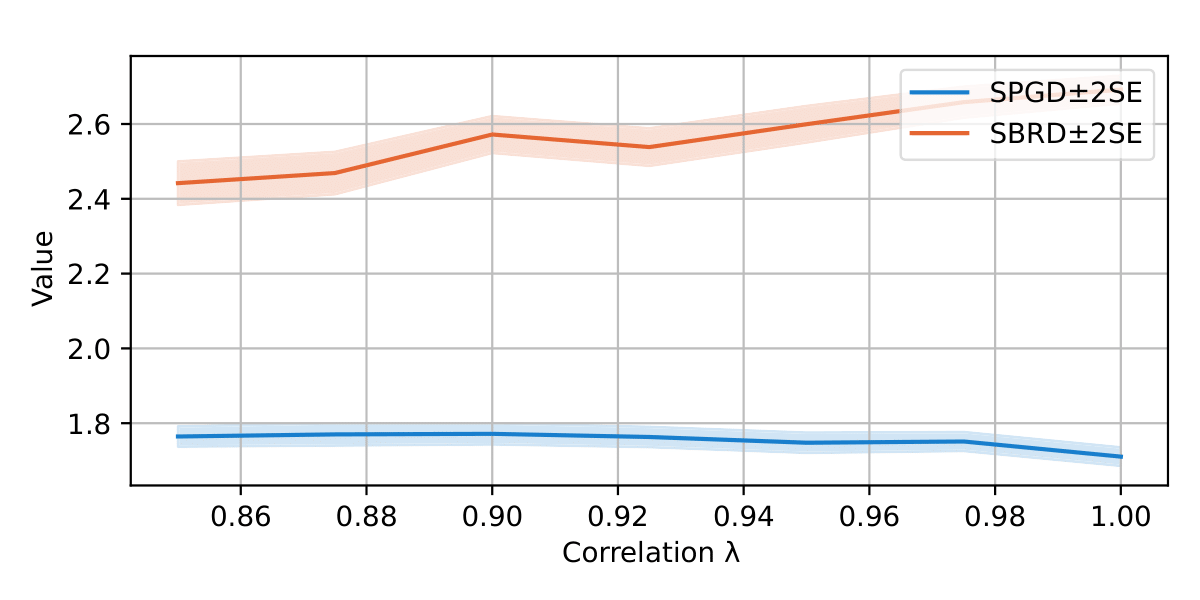}\label{S3ValAvg} }}%
    \caption{Comparison of SPGD and SBRD in a three-player $50$-actions game. $1000$ samples were drawn. Runtime: $80$ minutes.}%
    \label{fig:comparison_2}%
\end{figure}

We now justify our claim that SBRD provides a viable alternative to SPGD when the correlation is high, especially in online settings. We do this by examining the trade-off between convergence speed and final payoff. As shown in Section~\ref{exp_comparison}, SBRD typically reaches slightly lower equilibrium payoffs than SPGD. However, Figure~\ref{S3Steps} demonstrates that SBRD converges in significantly fewer steps, allowing agents to begin benefitting from equilibrium payoffs much earlier.

When comparing the average payoff of SPGD along its learning trajectory with the final payoff obtained by SBRD (as shown in Figure~\ref{S3ValAvg}), we find that SPGD accumulates substantially lower rewards during training. This suggests that in online settings, or in environments where short to medium time horizons are critical, SBRD may be the preferable choice.

In the next section we show that these differences become even more pronounced when the number of actions increases.

The experiment shown in Figure~\ref{fig:comparison_2} ran in $80$ minutes.

\subsection{Robustness to Number of Actions for Section ~\ref{exp_comparison}}\label{appendix_Robustness_comparison}
\begin{figure}%
    \centering
    \subfloat[\centering Time to convergence in SBRD and SPGD.]{{\includegraphics[width=6.7cm]{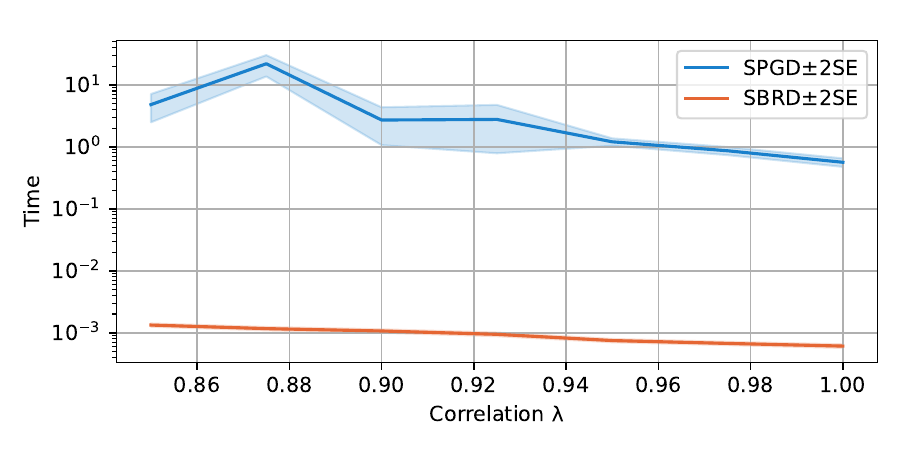} }}%
    \subfloat[\centering Final value of equilibrium.]{{\includegraphics[width=6.7cm]{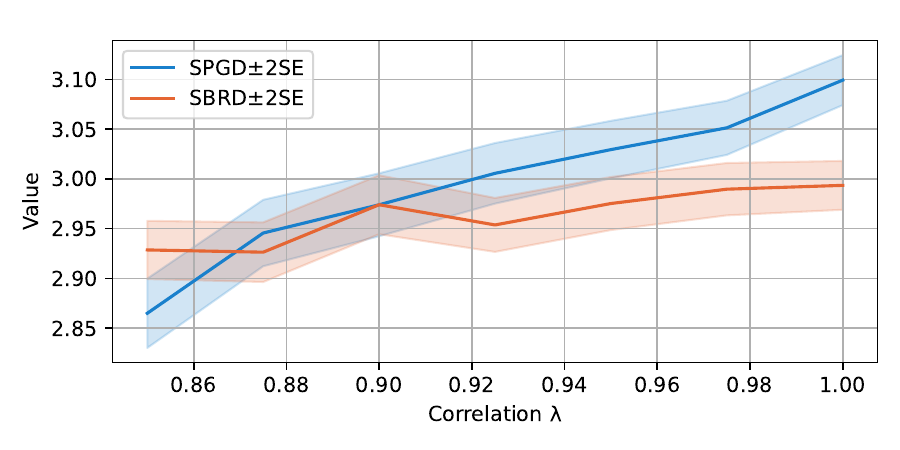} }}%
    \quad
    \subfloat[\centering Steps to convergence in SBRD and SPGD.]{{\includegraphics[width=6.7cm]{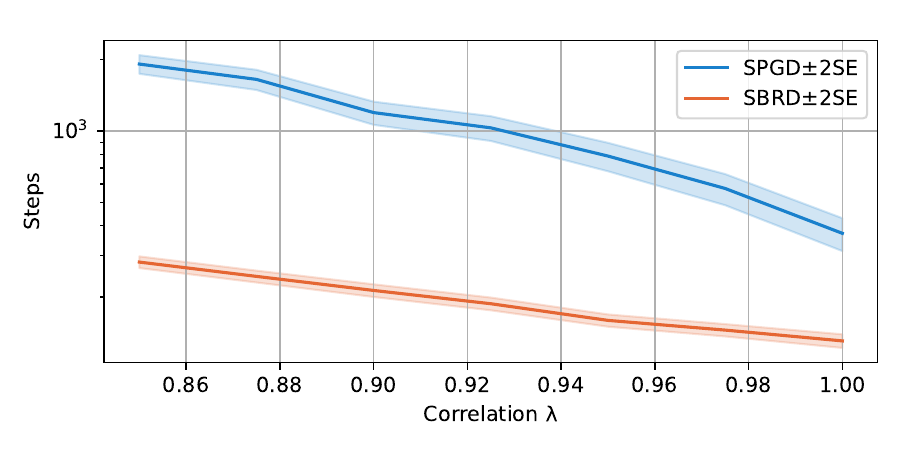} }}%
    \subfloat[\centering SBRD equilibrium value vs SPGD average value.]{{\includegraphics[width=6.7cm]{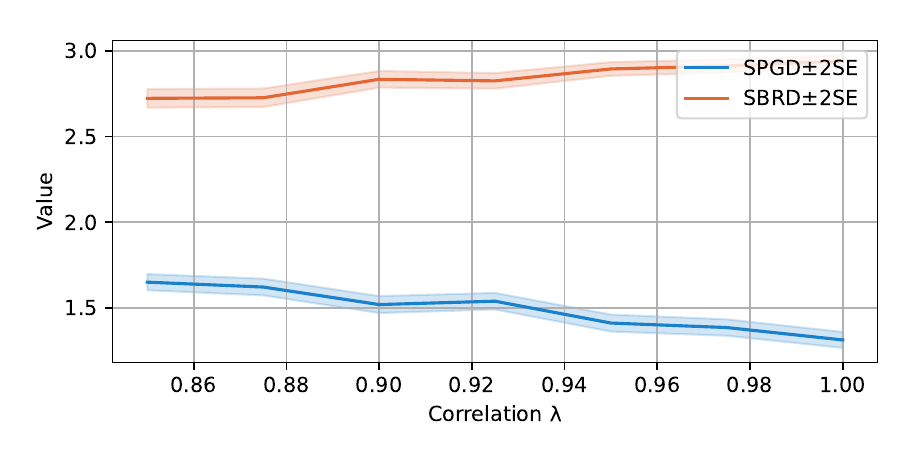} }}%
    \caption{Comparison of SPGD and SBRD in a three-player $100$-actions game. $1000$ samples were drawn. Runtime: $585$ minutes.}%
    \label{fig:comparison_more_actions}%
\end{figure}
Finally, we present further evidence that highly correlated three-player random games with $100$ actions exhibit behaviour consistent with the $50$-action case discussed earlier. The findings of this section can all be found in Figure~\ref{fig:comparison_more_actions}.

In particular, the findings show that SBRD converges to a Nash equilibrium significantly faster than SPGD (three to four orders of magnitude faster). This confirms the scalability of SBRD's performance as the size of the action space increases.

Moreover, we can see that the payoffs attained by both algorithms at equilibrium are closely comparable in magnitude. Notably, for relatively lower values of correlation, SBRD on average achieves better equilibrium values than SPGD. This highlights that SBRD's faster convergence does not come at a substantial cost in reward quality.

As with the $50$-action experiments, we also find that SBRD requires far fewer steps to reach convergence. This reinforces our claim that SBRD is particularly well suited for online or time-sensitive environments. In such settings, agents often benefit more from earlier access to high-value strategies than from long-term optimality alone. Since the average payoff collected by SPGD along its learning trajectory is consistently lower than the payoff achieved at equilibrium by SBRD, the latter emerges as a  competitive alternative in scenarios with limited time horizons.

The experiment shown in Figure~\ref{fig:comparison_more_actions} ran in $585$ minutes.

\end{document}